\documentclass[preprint]{revtex4-1}
\usepackage{graphicx}
\usepackage{subfig}
\usepackage{color}
\usepackage{array}
\usepackage{amsmath}
\newcolumntype{M}{>{\centering\arraybackslash}m{\dimexpr.25\linewidth-2\tabcolsep}}

\begin{document}

\title{Maximal Random Packing of Spheres in Menger Sponge}

\author{Micha\l{} Cie\'sla$^{1}$}
 \email{michal.ciesla@uj.edu.pl}
\author{Jakub Barbasz$^{1,2}$}
 \email{ncbarbas@cyf-kr.edu.pl}

\affiliation{
$^1$ M. Smoluchowski Institute of Physics, Jagiellonian University, 30-059 Kraków, Reymonta 4, Poland. \\
$^2$ Institute of Catalysis and Surface Chemistry, Polish Academy of Sciences, 30-239 Kraków, Niezapominajek 8, Poland.
}
\date{\today}

\begin{abstract}
Random packing of spheres inside fractal collectors of dimension $2<d<3$ is studied numerically using Random Sequential Adsorption algorithm. The paper focuses mainly on the measurement of maximal random packing ratio. Additionally, scaling properties of density autocorrelations in the obtained packing are analyzed. The RSA kinetics coefficients are also measured. Obtained results allow to test phenomenological relation between maximal random coverage ratio and collector dimension. Additionally, performed simulations together with previously obtained results confirm that, in general, the known dimensional relations are obeyed by systems having non-integer dimension at least for $d<3$.

\end{abstract}

\maketitle

\section{Introduction}
Hard spheres packing has a reach research history, especially in maths but also in natural science, due to it's potential application. Attention has been mostly paid to maximal  packing as it has many important implications, for example in crystallography. The fundamental property defining packing is its density \cite{bib:Conway1998, bib:Rogers1964}:
\begin{equation}
\theta_{max}(d) = \rho \cdot v(d),
\end{equation}
where $d$ is dimension of space, $\rho$ is a number of spheres in a unit volume, and
\begin{equation}
v(d) = \frac{ \pi^{d/2} }{ \Gamma (1+d/2) } \,\, r_0^d 
\label{eq:vd}
\end{equation}
is a volume of $d$ dimensional sphere having radius $r_0$. $\Gamma(x)$ is the Euler gamma function.
\par
The problem of maximal spheres packing is trivial in one dimension, simple in two dimensions and only recently has been proved in three dimensional space \cite{bib:Hales2006}. For higher dimensions, it still remains open. 
\par
Presented work, however, focuses on maximal random packing. It is derived from colloids study, namely, modelling of irreversible adsorption processes, in which molecules are randomly placed on a collector surface. Once adsorbed, a molecule retains its position on a surface. Such adsorption layer grows till there is available uncovered place large enough to place subsequent molecules. As the simplest and most commonly used model of adsorbed molecule is sphere, the properties of adsorption layers are directly connected with random maximal packing of spheres. Here, the only analytically solved case is a one-dimensional problem, also known also as car parking problem, for which $\theta_{max}(1)=0.748...$\cite{bib:Renyi1963}. Solutions in higher dimensions were found only numerically, however, it is worth to notice that there is a number of analytically solvable models for lattice-like collectors e.g.~\cite{bib:Privman2004, bib:Fan1991}. For flat, homogeneous and continuous collectors, the most extensively studied case is a two-dimensional problem due to its potential application in chemistry and material science \cite{bib:Feder1980, bib:Swendsen1981, bib:Privman1991, bib:Torquato2006, bib:Torquato2002}. The analysis of hard spheres random packing for $2 \le d \le 6$, nicely reviewed in \cite{bib:Torquato2006}, provides phenomenological relation, further slightly corrected later in \cite{bib:Ciesla2012b}, between  maximal random coverage ratio and collector dimension:
\begin{equation}
\theta_{max}(d) = \frac{1 + c_1 d + c_2 d^2}{2^d},
\label{eq:fit}
\end{equation}
with $c_1=0.416$ and $c_2=0.086$. 
There are also some characteristics, which do not depend qualitatively on dimension. For example, the two-point density correlation function for maximal random coverages is known to have super-exponential decay with growing distance \cite{bib:Bonnier1994}
\begin{equation}
C(r) \sim \frac{1}{ \Gamma(r) } \left( \frac{2}{\ln \left( r/(2r_0) -1 \right)} \right)^{(r/2r_0)-1} \mbox{ for } r \to \infty, 
\label{eq:corfast}
\end{equation}
and logarithmic singularity when particles are in touch \cite{bib:Swendsen1981, bib:Privman1991}
\begin{equation}
C(r) \sim \ln \left( \frac{r}{2r_0} -1 \right) \mbox{ for } r \to (2r_0)^+ \mbox{ and } \theta \to \theta_{max}.
\label{eq:corlog}
\end{equation}
\par
Aim of present work is to check validity of the above relations for adsorption on fractal collectors with structure similar to the Menger sponge. We decided to focus on this kind of collectors because its structure remains zeolites, which were extensively studied in the context of adsorption, e.g.\cite{bib:Fuchs2001}.
\par
Adsorption experiments using fractal-like collectors were made by Kinge et al.\ \cite{bib:Kinge2008}, though preparation of such surfaces is quite complicated. On the other hand, there are number of natural porous media when adsorption plays an important role there \cite{bib:Pfeifer1983, bib:Avnir1983}. For example,  fractal-like structure of corals helps them to catch plankton effectively \cite{bib:Basillais1998}. Adsorption on fractal collectors might also be applied in environmental protection in designing effective filters \cite{bib:Khasanov1991}. Other studies on fractal collectors investigate diffusion properties \cite{bib:Nazzarro1996, bib:Loscar2003} or adsorption on rough surfaces \cite{bib:Cole1986, bib:Pfeifer1989}. On the other hand, fractal structures were also observed as a result of adsorption \cite{bib:Brilliantov1998}. However, the only similar study concerning adsorption on fractal collectors is limited to collectors of $D<2$ \cite{bib:Ciesla2012b}.
\par
The most common method of obtaining numerically the maximal random coverages uses Random Sequential Adsorption (RSA) algorithm \cite{bib:Feder1980} described in detail in Sec.\ref{sec:Model}. The kinetics of RSA simulation has also been of our interest because it affects $\theta_{max}(d)$ estimation. Moreover, it was observed that for sufficiently long simulation time, the coverage ratio $\theta(t)$ scales with collector's dimension \cite{bib:Feder1980, bib:Swendsen1981}:
\begin{equation}
\theta_{max} - \theta(t) \sim t^{-1/d}.
\label{eq:feder}
\end{equation}
Here $\theta_{max} \equiv \theta(t \to \infty)$. The above relation is all the more important as it is also valid for other molecules, more complex than simple spheres \cite{bib:Ciesla2012a, bib:Adamczyk2010, bib:Adamczyk2011}.
\section{Model}
\label{sec:Model}
Maximal random coverages are generated using RSA algorithm, which has been successfully applied to study colloids. It is based on independent, repeated attempts of adding sphere to a covering film. The numerical procedure is carried out in the following steps:
\begin{description}
\item[i] a virtual sphere is created with its centre position on a collector chosen randomly according to the uniform probability distribution;
\item[ii] an overlapping test is performed for previously adsorbed nearest neighbors of the virtual particle. The test checks if surface-to-surface distance between spheres is greater than zero;
\item[iii] if there is no overlap the virtual particle is irreversibly adsorbed and added to an existing covering layer. Its position does not change during further calculations;
\item[iiii] if there is an overlap the virtual sphere is removed and abandoned.
\end{description}
Attempts are repeated iteratively. Their number is typically expressed in a~dimensionless time units:
\begin{equation}
t_0 = N\frac{S_\text{D}}{S_\text{C}},
\end{equation}
where $N$ is a number of attempts and $S_\text{D}$ denotes a volume of a single sphere projection on a collector (note that only the center of a sphere has to be on collector) and $S_\text{C}$ is a collector volume, assuming its integer dimension. Here, in $D<3$ the volume of any collector is formally zero. On the other hand, any given value of $S_\text{C}$ is only a scale factor of $t_0$. Therefore, for the purposes of our study, we assumed $S_\text{C} = (100 \cdot r_0)^3$, which is the volume of an overlapping cube. Collectors were modelled as a successive iteration of a given fractal. They were constructed starting with a single cube, then dividing it into 27 subcubes and finally removing some of them. Then the procedure was repeated iteratively for each of the remaining subcubes. The real fractal is achieved after an infinite number of iterations. Its dimension is given by relation
\begin{equation}
d = \frac{\ln(p)}{\ln(s)},
\end{equation}
where $p$ is a number of remaining subcubes in each iteration and $s$ is a change of scale. In our case, $s=3$. For example for Menger sponge $p=20$ so $D=\ln(20) / \ln(3) \approx 2.727$. With each of the 27 subcubes marked with its Cartesian coordinates, the fractal can be fully described by coordinates of removed or retained subcubes. For convenience, basic properties of fractal collectors used in this study were collected in Tab.\ref{tab:fractals}.
\begin{table}[ht]
  \centering
  \begin{tabular}{MMM}
  removed subcubes coordinates & fractal dimension d & 1$^{st}$ iteration of a~collector\\
  \hline \vspace{1cm}
  all but 113, 131, 222, 311, 333  &  1.4650  & \includegraphics[width=3cm]{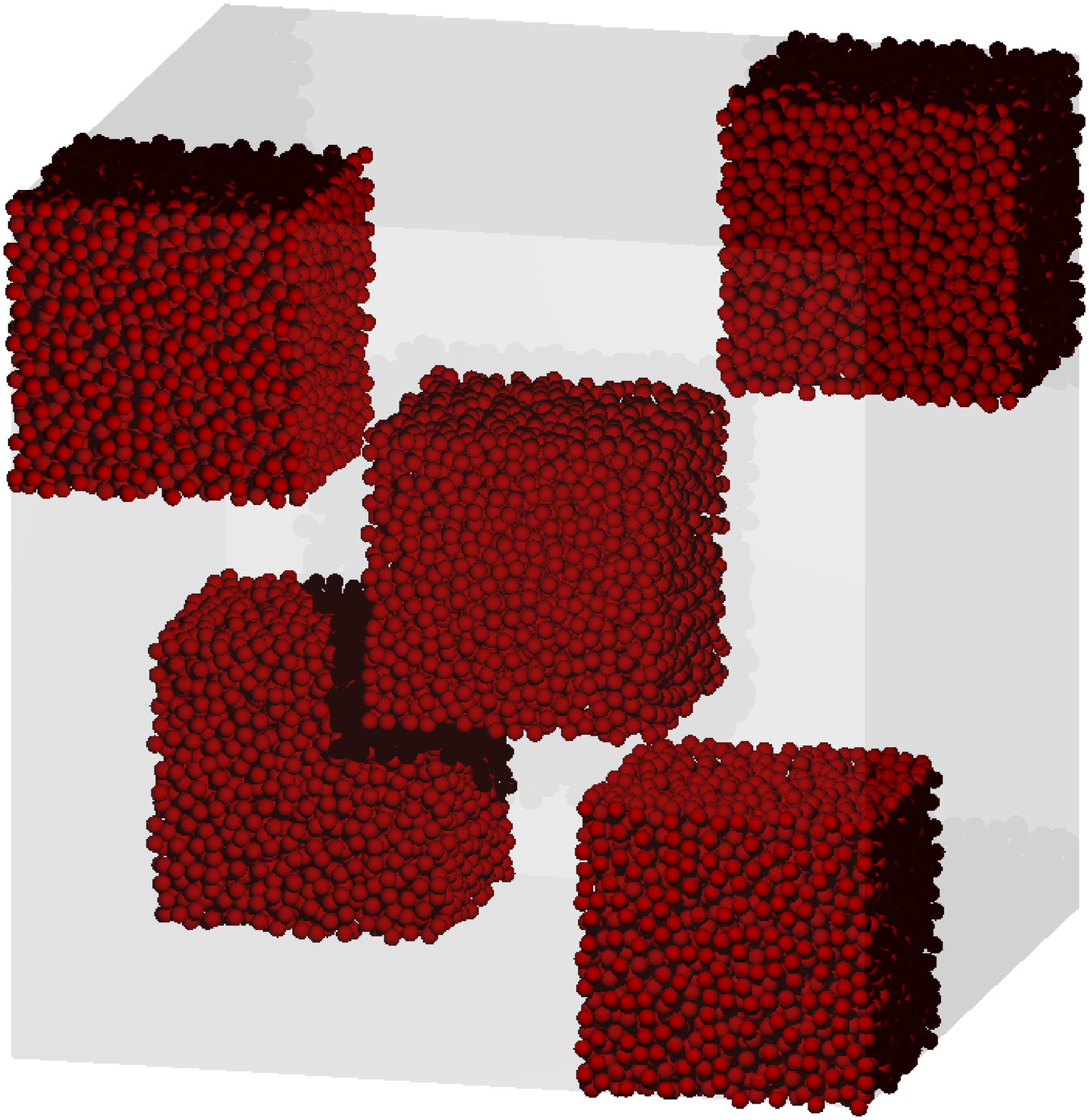} \\
  all but 111, 113, 131, 133, 222, 311, 313, 331, 333 &  2.0 & \includegraphics[width=3cm]{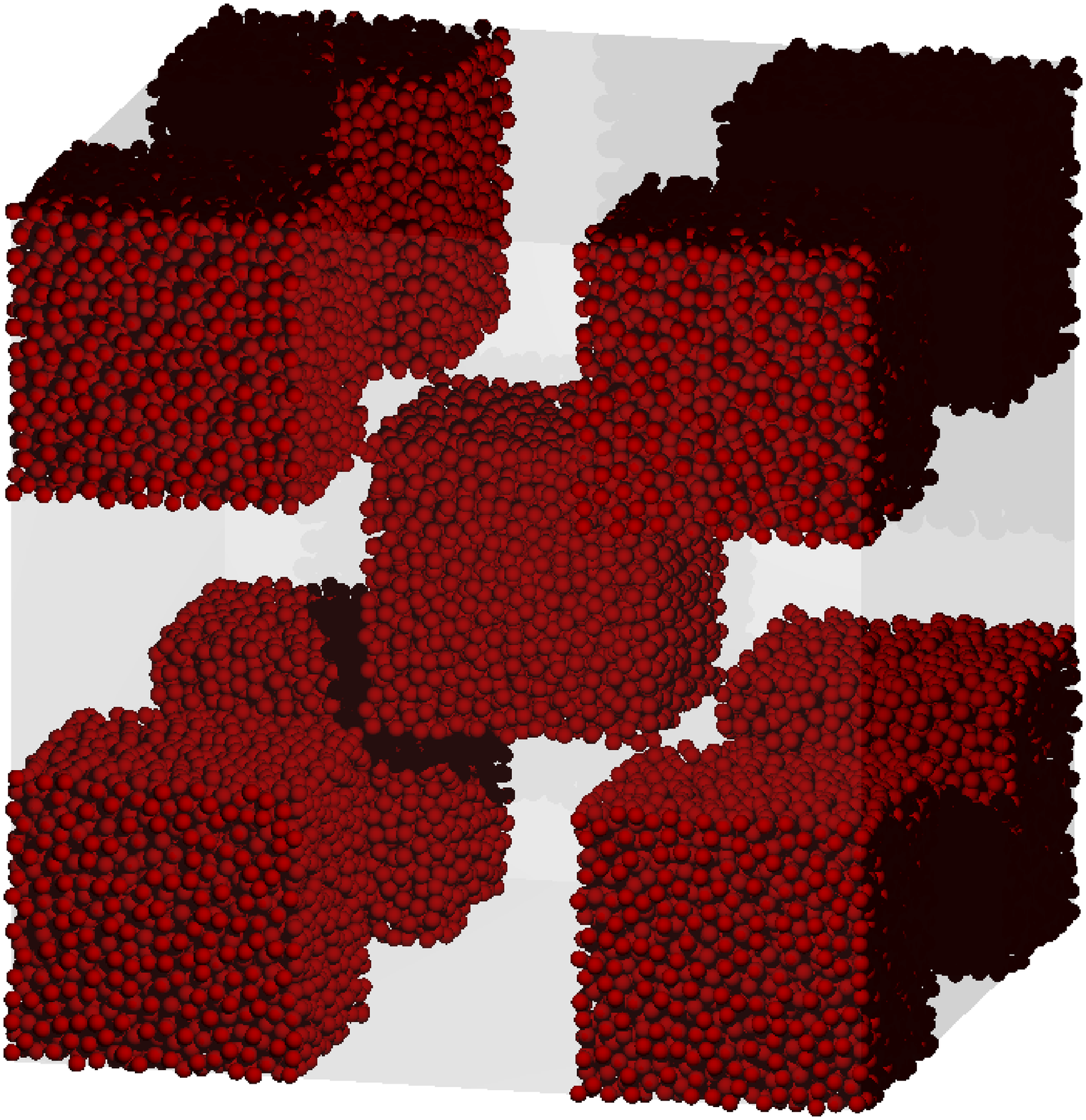} \\
  112, 121, 123, 132, 211, 213, 222, 231, 233, 311, 313, 331, 333 & 2.4022 & \includegraphics[width=3cm]{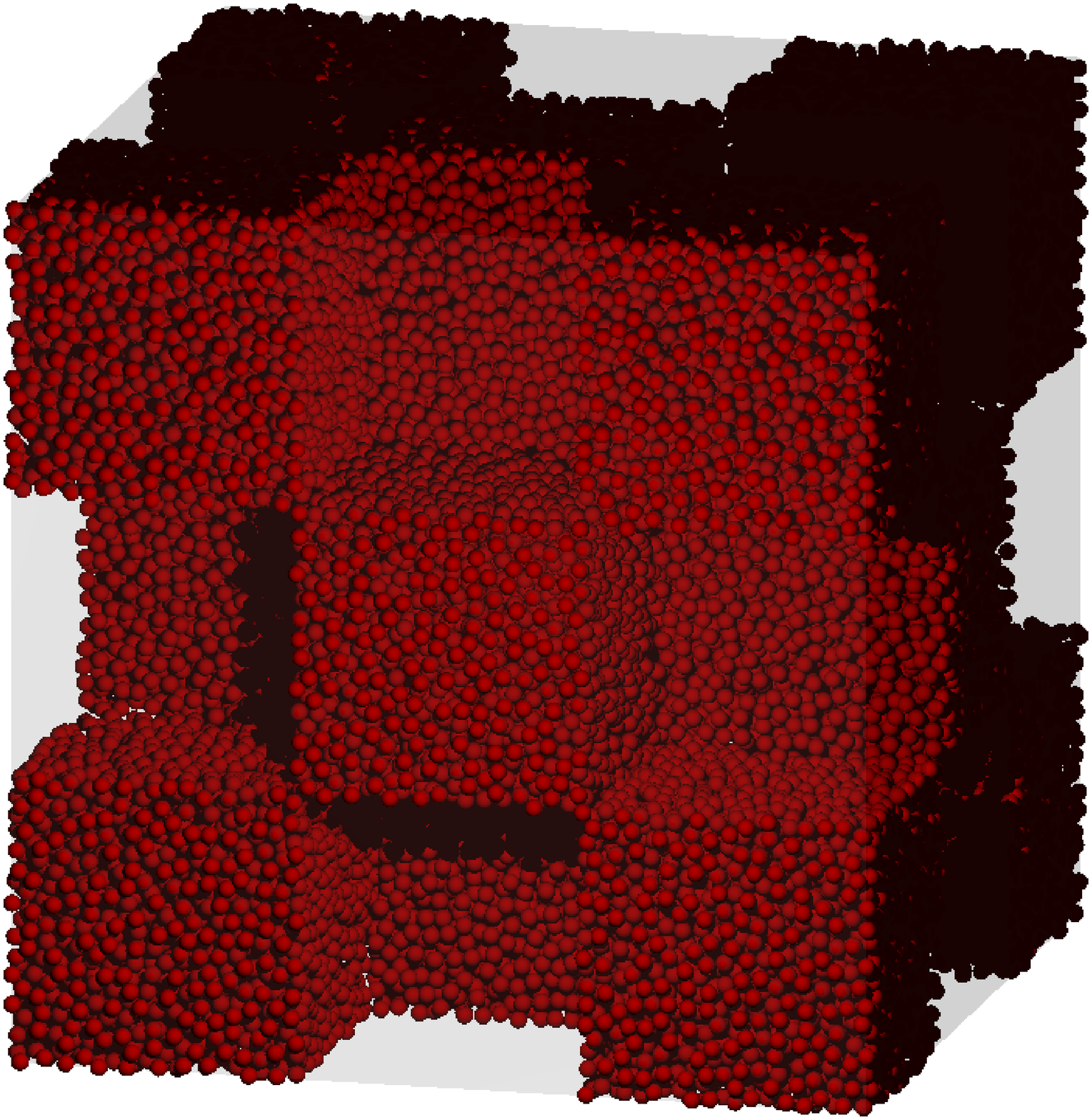} \\
  122, 211, 212, 213, 221, 223, 231, 232, 233, 322 & 2.5789  & \includegraphics[width=3cm]{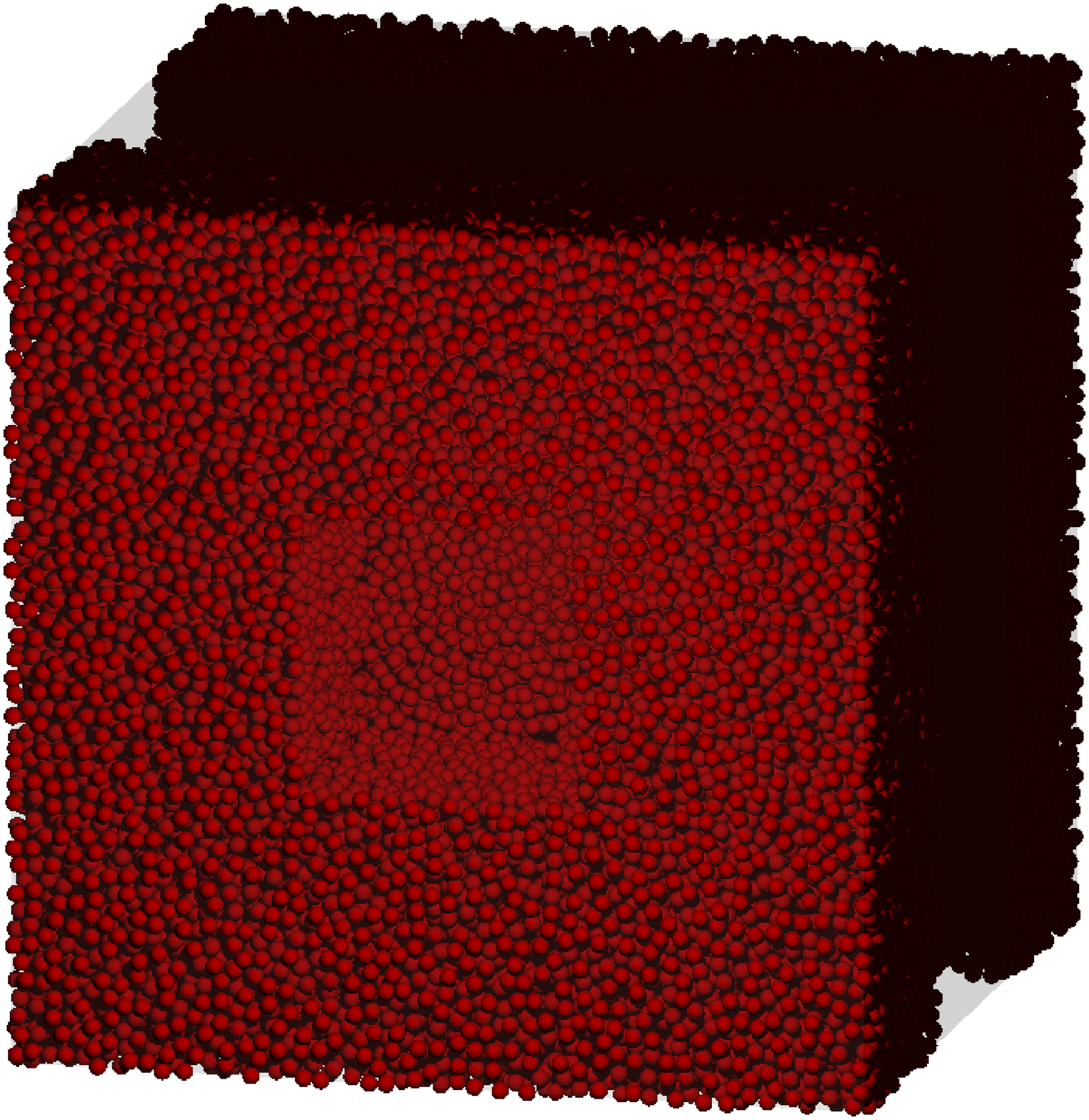} \\
  122, 212, 221, 222, 223, 232, 322 & 2.7268 & \includegraphics[width=3cm]{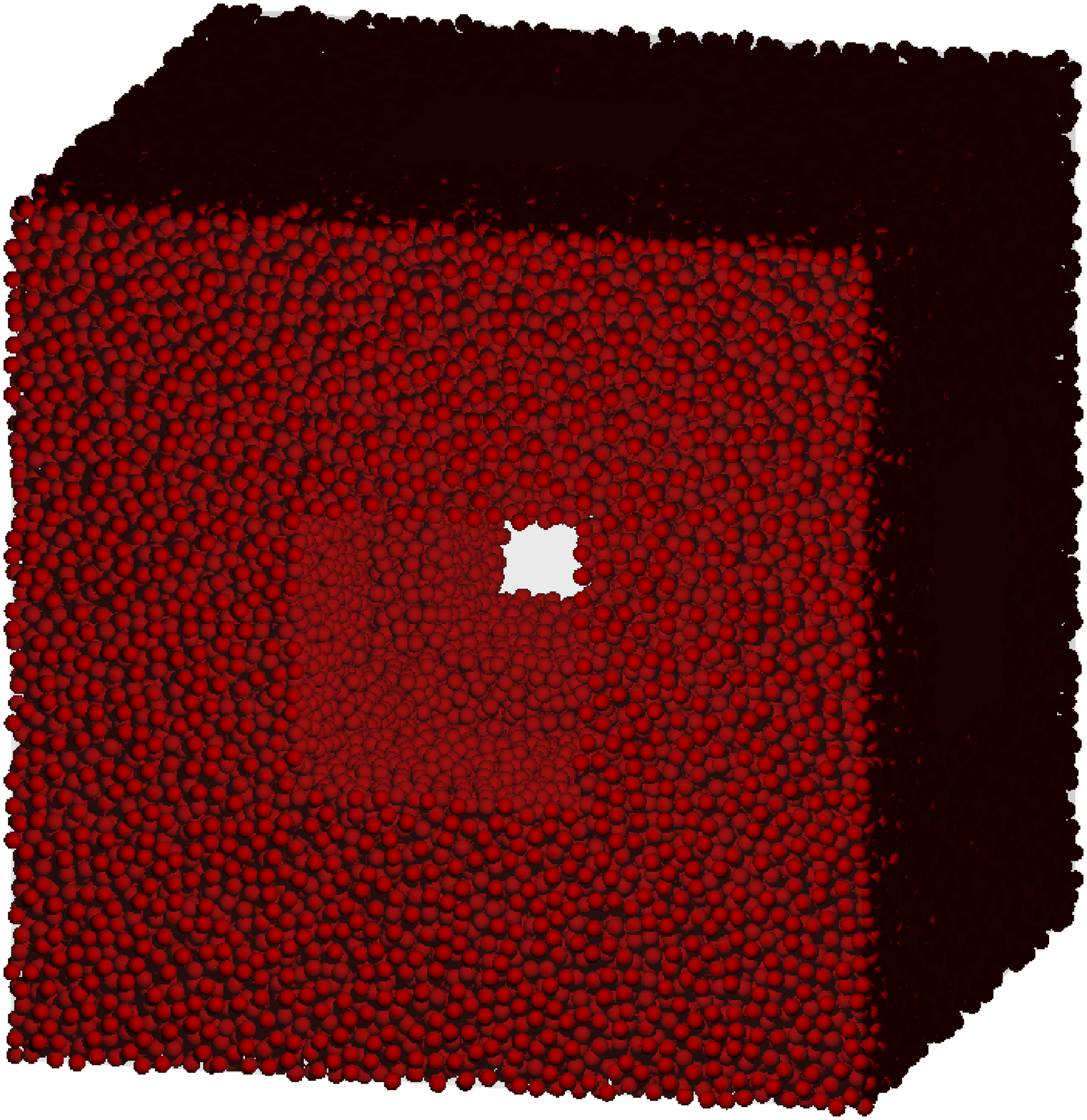} \\
  211, 213, 231, 233 & 2.8540 & \includegraphics[width=3cm]{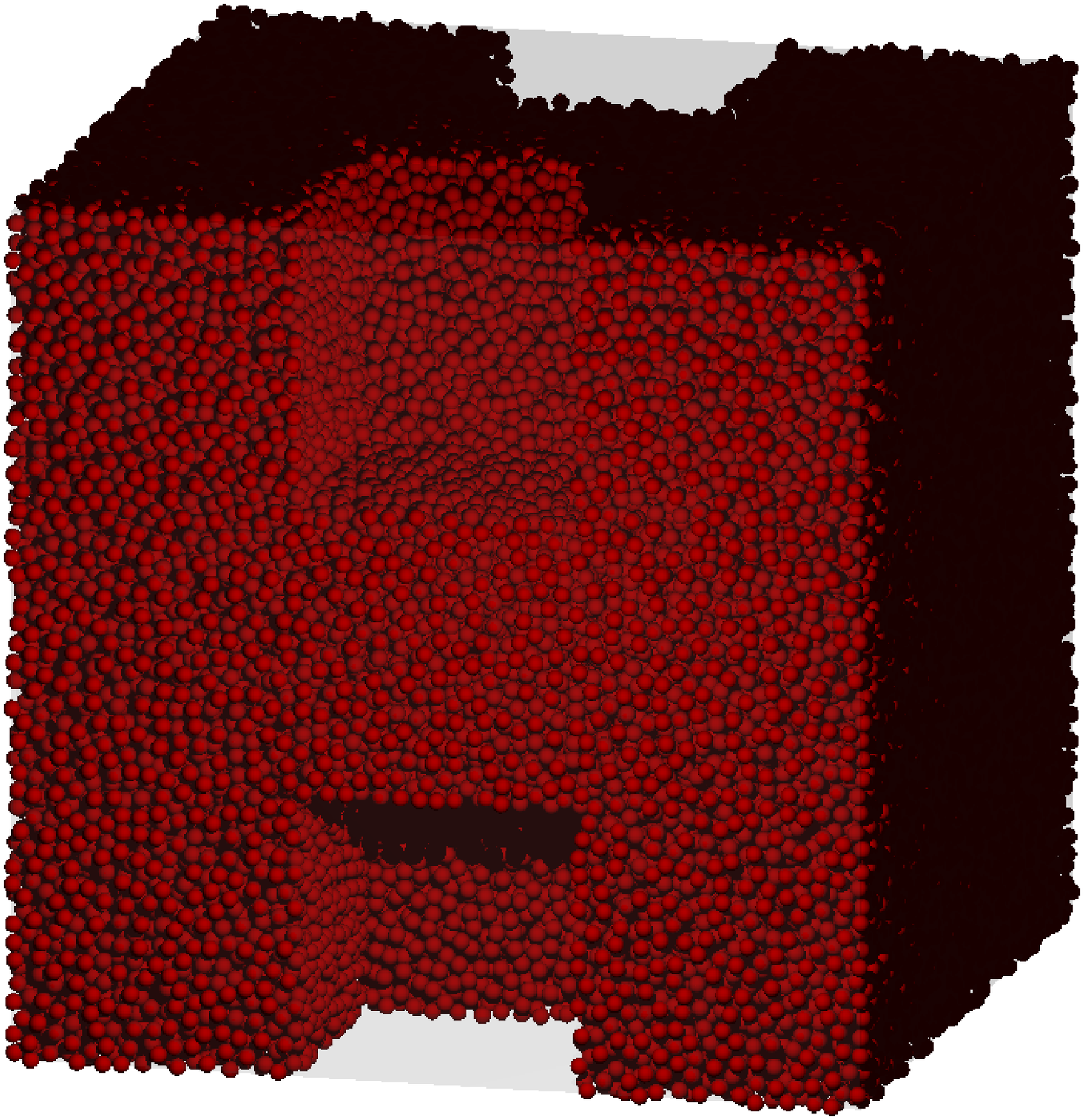} \\
  222 & 2.9656 & \includegraphics[width=3cm]{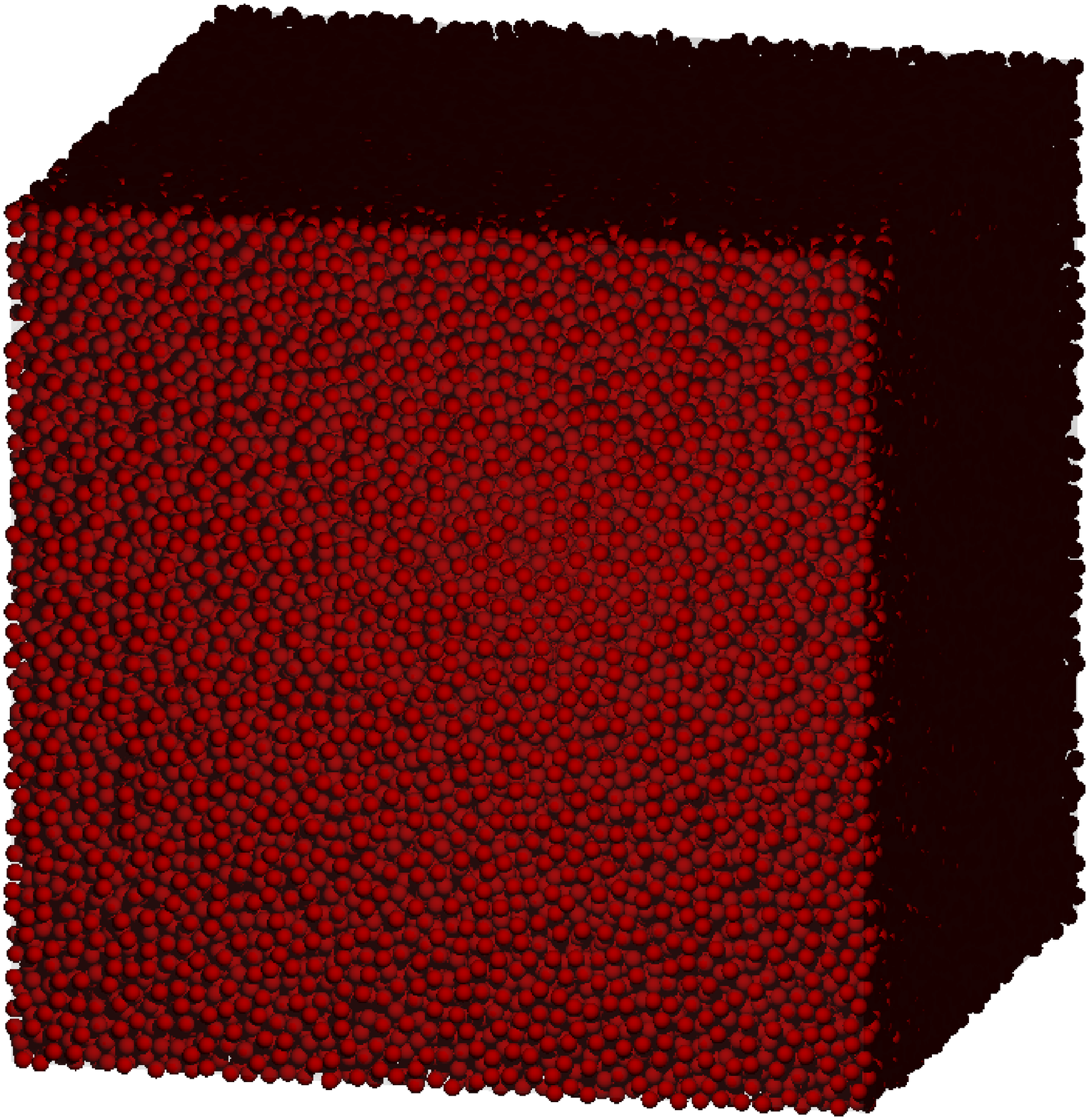} \\  
  \hline
  \end{tabular}
  \caption{Definitions and fractal dimensions of collectors used for adsorption modelling.} 
  \label{tab:fractals}
\end{table}
For each collector type $100$ independent numerical experiments have been performed. Each of them last $10^4 t_0$.
\par 
Typically coverage ratio is estimated by a number of adsorbed molecules. However, this approximation can be misleading when collector dimension differs from adsorbate dimension. Therefore, by analogy with integer dimension cases, covered area is limited to a cross-section of collector and adsorbate sphere. For this reason, the coverage ratio of a given adsorption layer is determined using random sampling of collector points and checking whether they are covered by any sphere adsorbed earlier:
\begin{equation}  
\label{eq:qt}
\theta(t) = \frac{n_c(t)}{n},
\end{equation}
where $n_c(t)$ is a number of covered points after a dimensionless time $t$; $n$ denotes total number of sampled points. Here $n=10^6$, which provides statistical error at the level of $0.1\%$.
\section{Results and Discussion}
\begin{figure}[htb]
\vspace{1cm}
\centerline{%
\includegraphics[width=8cm]{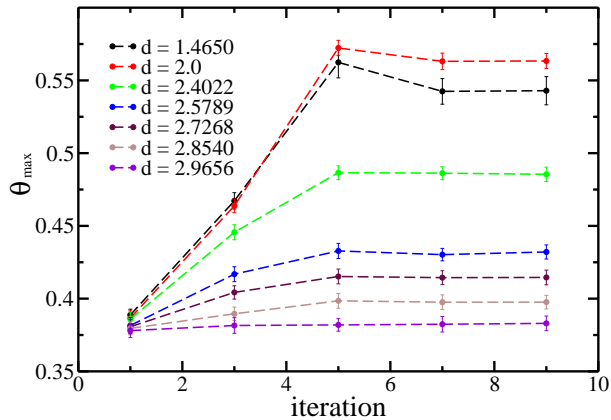}}
\caption{The dependence of maximal random coverage ratio on fractal iteration for all studied fractals.}
\label{fig:iterations}
\end{figure}

In performed simulations a real fractal was approximated by a its finite iteration. To check the level of possible systematic error introduced by such approximation we plotted maximal random coverage as a function of fractal iteration. Results presented in Fig.\ref{fig:iterations} show that measured values stabilize around $7^{th}$ iteration.
Therefore, for the purpose of this study ,we assumed that results obtained for $9^{th}$ iteration math the true ones.
\subsection{Kinetics of the RSA}
\begin{figure}[htb]
\vspace{1cm}
\centerline{%
\includegraphics[width=8cm]{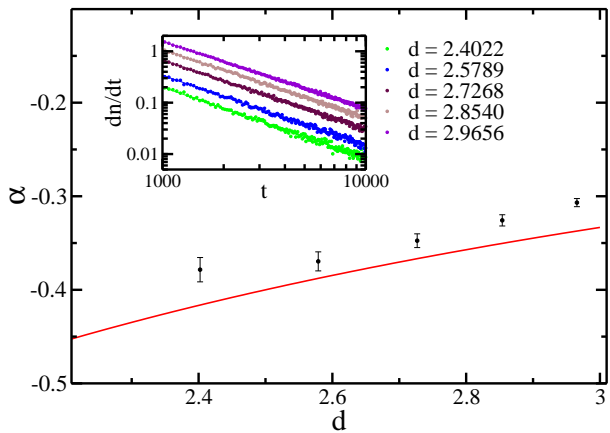}}
\caption{Feder's law validity. Dots represent exponents $\alpha$ whereas solid line is determined using Feder's law (\ref{eq:feder}). Inset shows power law dependence of the number of packed spheres on dimensionless time (\ref{qt}).}
\label{fig:p_t}
\end{figure}

The maximal random coverage is achieved after an infinite simulation time. Therefore to measure its value in finite simulations, the RSA kinetics model has to be used. The analysis of earlier numerical experiments shows that RSA for spheres in integer and fractal dimensions obeys the Feder's law (\ref{eq:feder}) \cite{bib:Torquato2006, bib:Ciesla2012b}. 
The most convenient way to determine exponent in (\ref{eq:feder}) is to find the time derivative of coverage ratio:
\begin{equation}
\frac{d \theta}{dt}(t) \sim t^{-1+\alpha},
\label{qt}
\end{equation}
where $\alpha = -1/d$. Results presented in Fig.\ref{fig:p_t} show that Feder's law slightly underestimates adsorption speed or equivalently slightly overestimates collector dimension. However, the difference would be practically imperceptible it the $\theta(t)$ dependence on dimensionless time was analyzed as in \cite{bib:Torquato2006, bib:Ciesla2012b}. For the sparsest collectors the relation (\ref{qt}) is not valid and, therefore, parameter $\alpha$ cannot be determined. This is analogous to the case of Cantor Dust and some Cantor sets discussed in \cite{bib:Ciesla2012b}.
\par
\begin{figure}[htb]
\vspace{1cm}
\centerline{%
\includegraphics[width=8cm]{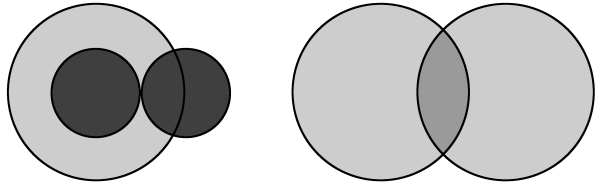}}
\caption{Left: excluded volume by single hypersphere ($C1$). Right: overlap of excluded volumes of two different particles.}
\label{fig:ev}
\end{figure}
\begin{figure}[htb]
\vspace{1cm}
\centerline{%
\includegraphics[width=8cm]{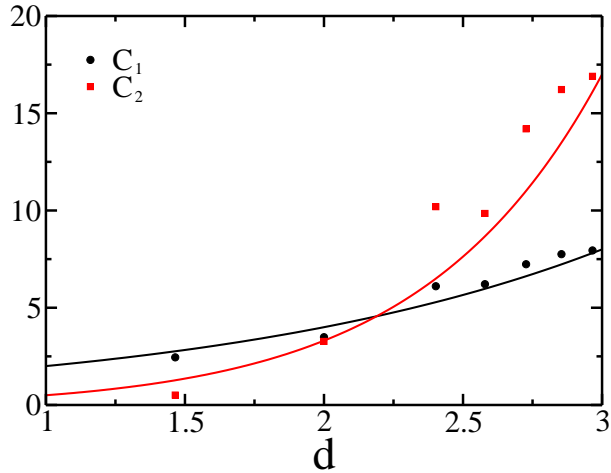}}
\caption{Dependence of blocking parameters $C_1$ and $C_2$ on collector dimension. Circles ($C_1$) and squares ($C_2$) represent obtained data while solid lines are analitical fits.}
\label{fig:c1c2}
\end{figure}

The adsorption speed $d \theta / dt$ may also be analyzed as a function of $\theta$; then, it is called Available Surface Function (ASF). Such approach is commonly used in material science and especially in low coverage limit, because it allows to determine viral coefficients of the equilibrium coverage state \cite{bib:Tarjus1991, bib:AdamczykBook}. 
\begin{equation}
\text{ASF}(\theta)= \frac{d \theta}{dt}(\theta) = 1 - C_1 \theta + C_2 \theta^2 + o(\theta^2),
\label{asf}
\end{equation}
where $C_1$ and $C_2$ are expansion constants and they are connected with volume blocked by particle and a complex of two particles, respectively. For example, in a low coverage limit, a single ball covers $v(d)$, but, from the point of view of another ball center, it blocks all the volume inside a sphere of $2r_0$ radius (see Fig.\ref{fig:ev}). 
Therefore, using (\ref{eq:vd})
\begin{equation}
 C_1(d) = \frac{(2r_0)^d}{r_0^d} = 2^d.
\end{equation}
For higher coverages the exclusion volumes start to overlap, which slightly reduces the total amount of blocked space. Therefore, by analogy with the reasoning presented in \cite{bib:AdamczykBook}, parameter $C_2$ is given by
\begin{equation}
C_2(d) = \frac{1}{2 v^2(d)} \int_{2r_0}^{4r_0}V_2(r, d)\cdot s(r, d) dr,
\end{equation}  
where
\begin{equation}
V_2(r, d) = 2\frac{\pi^\frac{d - 1}{2}(2r_0)^d}{\Gamma[(d + 1)/2]} \int_0^{\cos^{-1} \frac{r}{4r_0}} \sin^d t \,\,dt
\end{equation} 
is the overlapped volume and $s(r, d)$ is $(d-1)$-surface of $d$-sphere of radius $r$. 
Values of $C_1$ and $C_2$ measured in numerical experiments are shown in Fig.\ref{fig:c1c2}.
Values of $C_1$ fits well the theoretical approximation. In case of $C_2$ the match is slightly worse, however, numerical data are generally arranged along the analytically obtained values.
\subsection{The Maximal Random Coverage Ratio}
\begin{figure}[htb]
\vspace{1cm}
\centerline{%
\includegraphics[width=8cm]{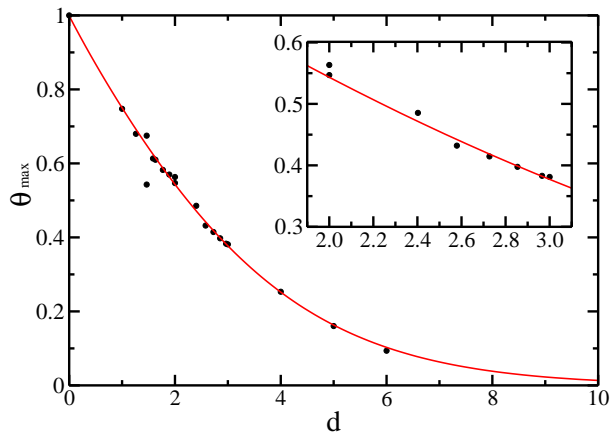}}
\caption{Maximal random coverage ratio in different dimensions. Dots represents data obtained here and also in \cite{bib:Torquato2006, bib:Ciesla2012b} and solid line is the fit (\ref{eq:fit}). Inset shows data for $d \in [2.0;\,\, 3.0]$ obtained in this study.}
\label{fig:qd}
\end{figure}

Despite the lack of analitical results for maximal random coverage ratio for $d>1$ there are several studies providing lower and upper limit of its value, e.g. \cite{bib:Ball1992}. The best relation describing dependence of saturated coverages on collector dimension, given by (\ref{eq:fit}) has been proposed in \cite{bib:Torquato2006, bib:Ciesla2012b} and fits our data, see Fig.\ref{fig:qd}.
For convenience, all numerical values obtained in this study and presented on above figures are put together in Tab.\ref{tab:kinetics}.
\par 
\begin{table}[ht]
  \centering
  \begin{tabular}{c c c c c}
  $d$ & $\theta_\text{max}$ & $\alpha$ & $C_1$ & $C_2$ \\	
  \hline
  $1.4650$ & $0.543 \pm 0.010$ &                 --              & 2.449 & 0.512 \\	
  $2.0$       & $0.563 \pm 0.005$ &                 --              & 3.485 & 3.268 \\
  $2.4022$ & $0.485 \pm 0.005$ & $-0.378 \pm 0.012$ & 6.105 & 10.194 \\
  $2.5789$ & $0.432 \pm 0.005$ & $-0.370 \pm 0.010$	& 6.211 & 9.849 \\
  $2.7268$ & $0.415 \pm 0.005$ & $-0.348 \pm 0.007$ & 7.239 & 14.197 \\
  $2.8540$	& $0.398 \pm 0.005$ & $-0.326 \pm 0.006$ & 7.754 & 16.218 \\
  $2.9656$ & $0.382 \pm 0.005$ & $-0.307 \pm 0.004$ & 7.945 & 16.899 \\
  \hline
  \end{tabular}
  \caption{Maximal random coverages ratio, RSA kinetics and coefficient $C_1$ and $C_2$ were (fitted for $\theta / \theta_\text{max}<0.3$) obtained from RSA simulation.}
  \label{tab:kinetics}
\end{table}
\subsection{Pair correlation function}
\begin{figure}[htb]
\vspace{0.1cm}
\centerline{%
\subfloat[]{\includegraphics[width=4cm]{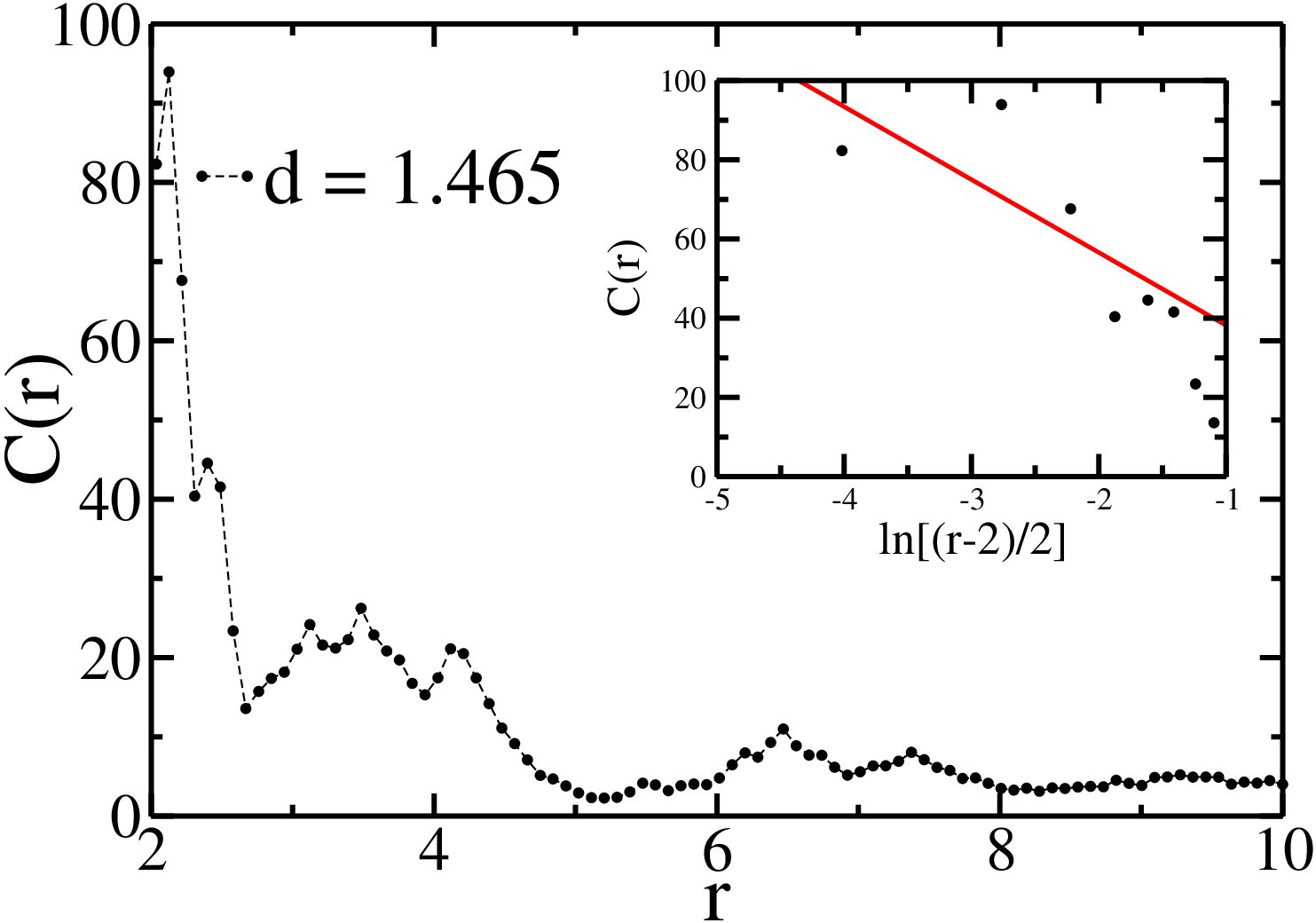}}
\hspace{1cm}
\subfloat[]{\includegraphics[width=4cm]{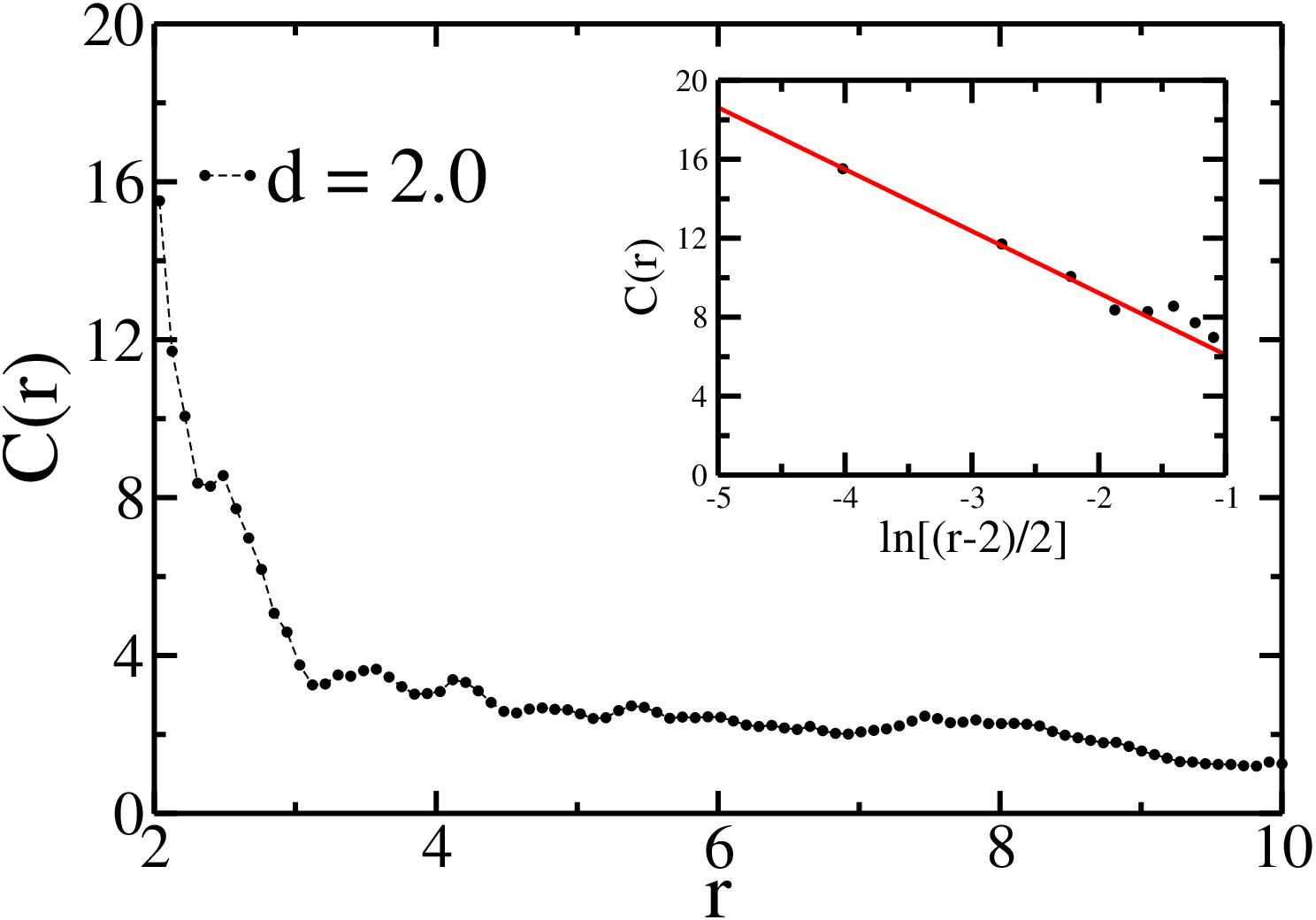}}
}
\vspace{0.1cm}
\centerline{%
\subfloat[]{\includegraphics[width=4cm]{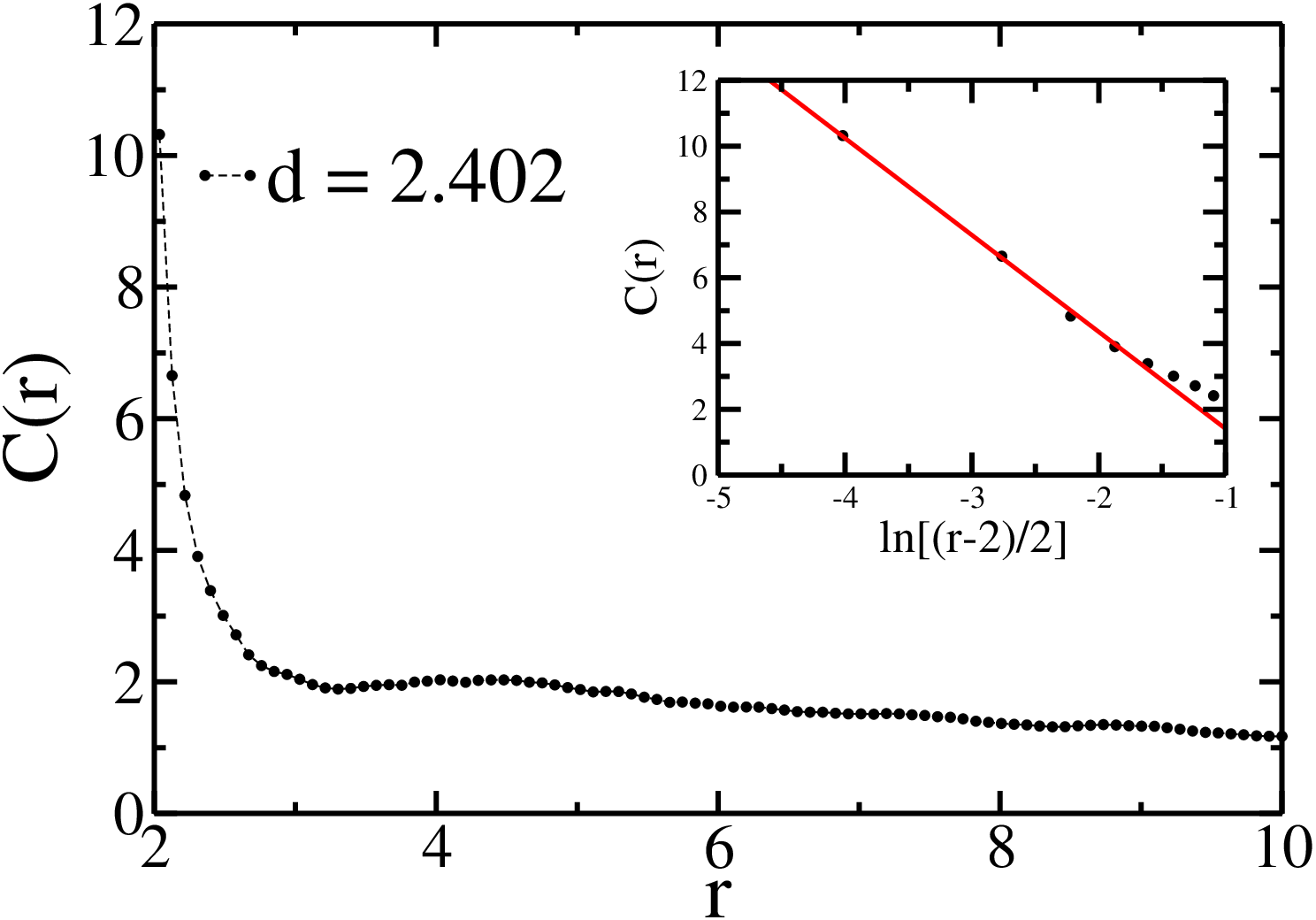}}
\hspace{1cm}
\subfloat[]{\includegraphics[width=4cm]{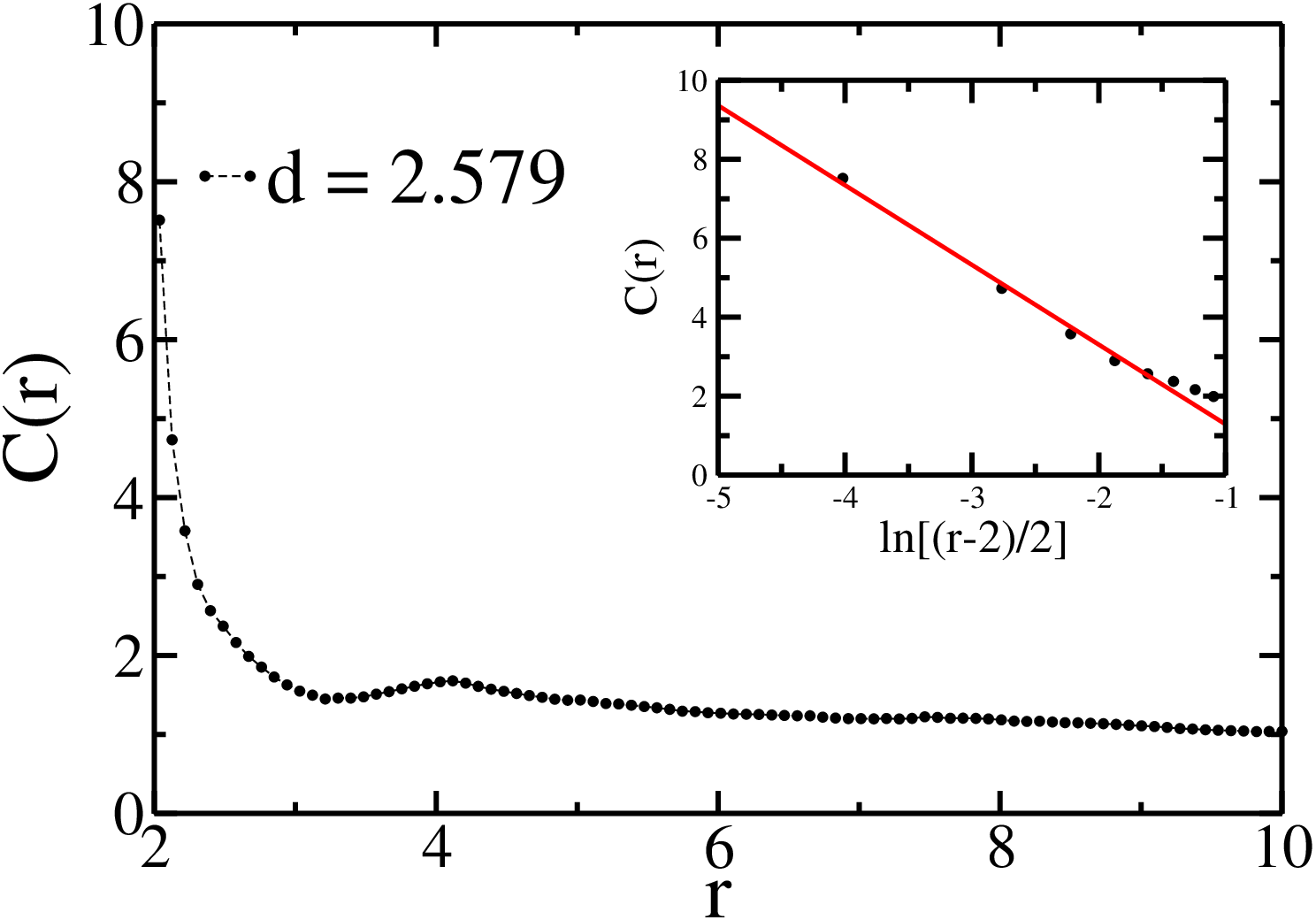}}
}
\vspace{0.1cm}
\centerline{%
\subfloat[]{\includegraphics[width=4cm]{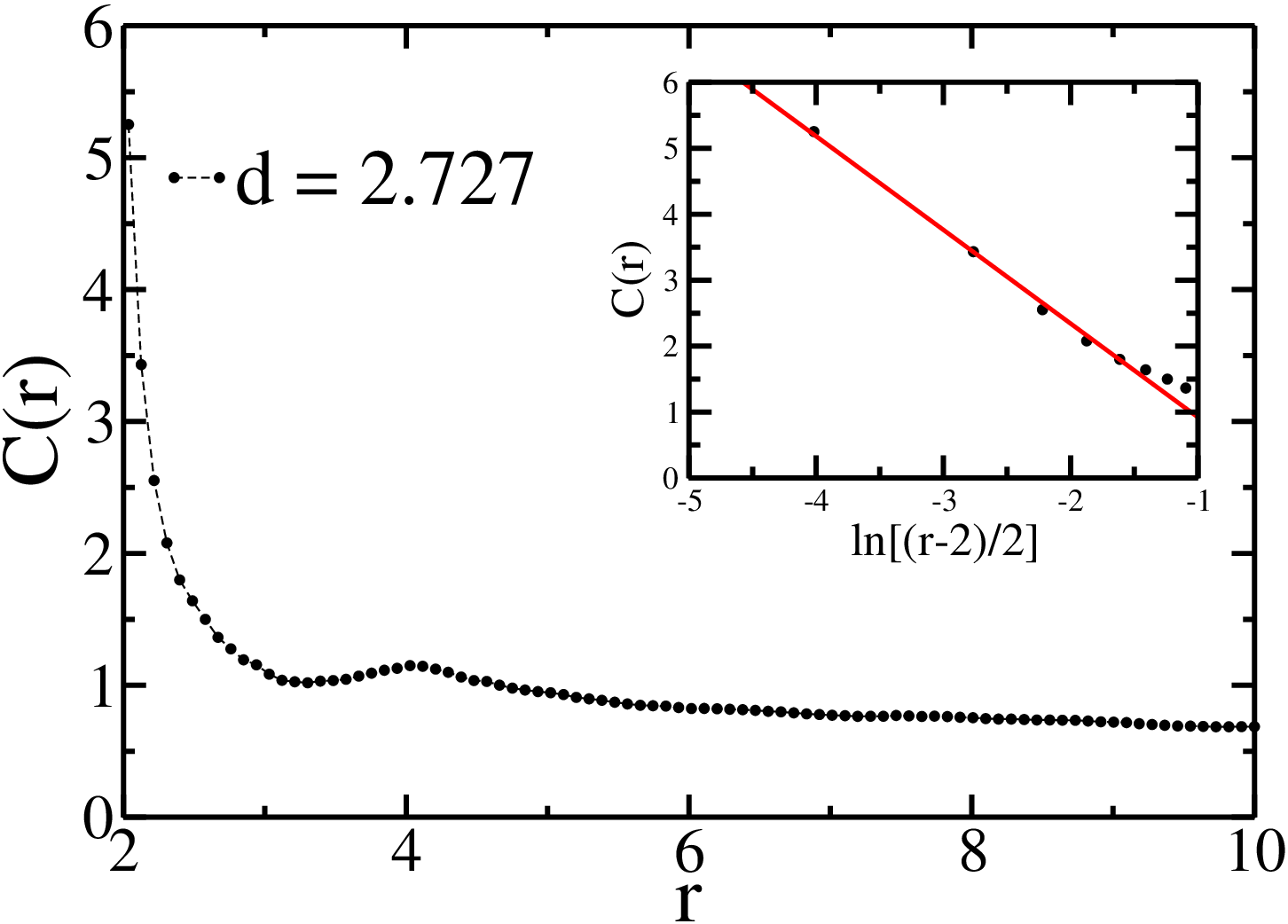}}
\hspace{1cm}
\subfloat[]{\includegraphics[width=4cm]{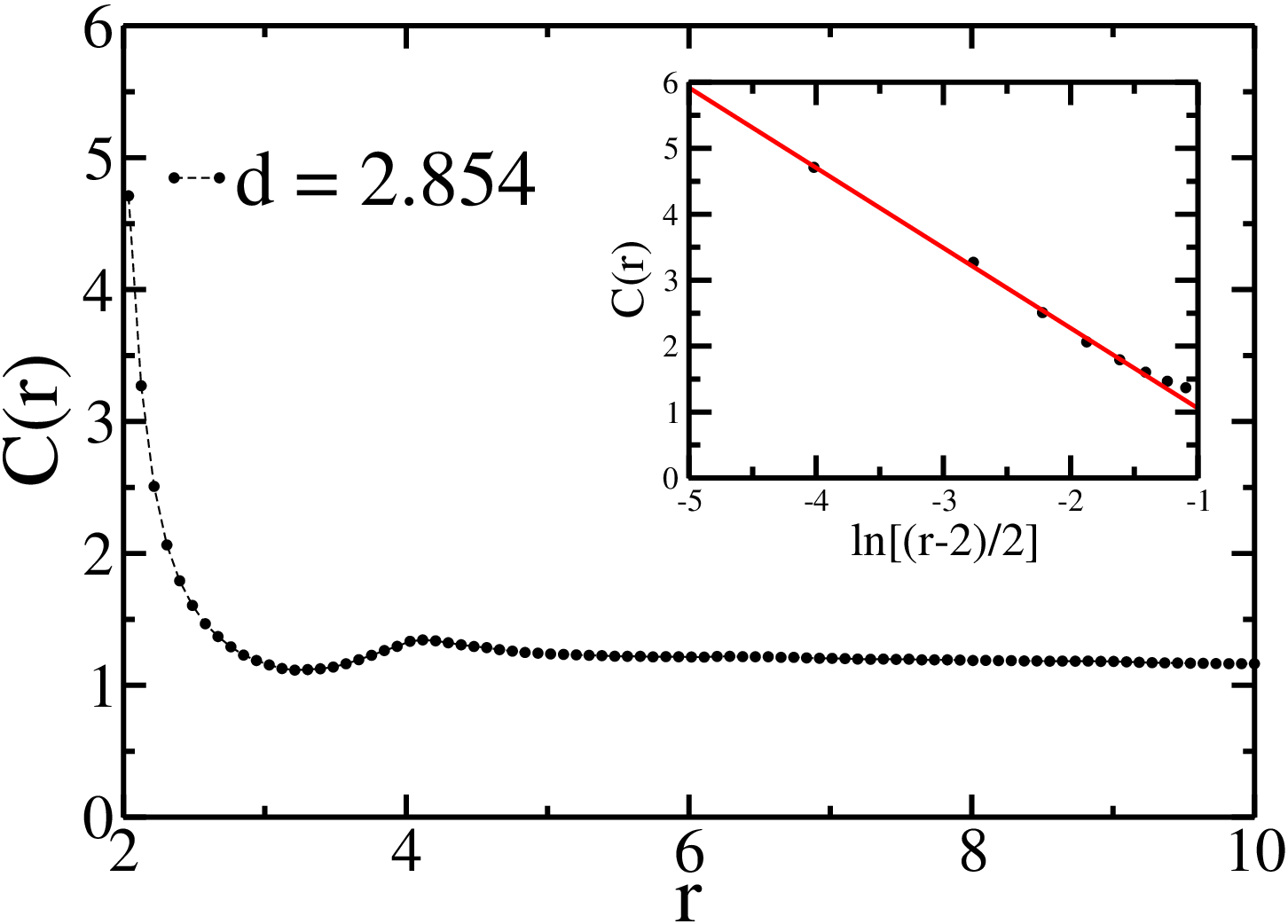}}
}
\vspace{0.1cm}
\centerline{%
\subfloat[]{\includegraphics[width=4cm]{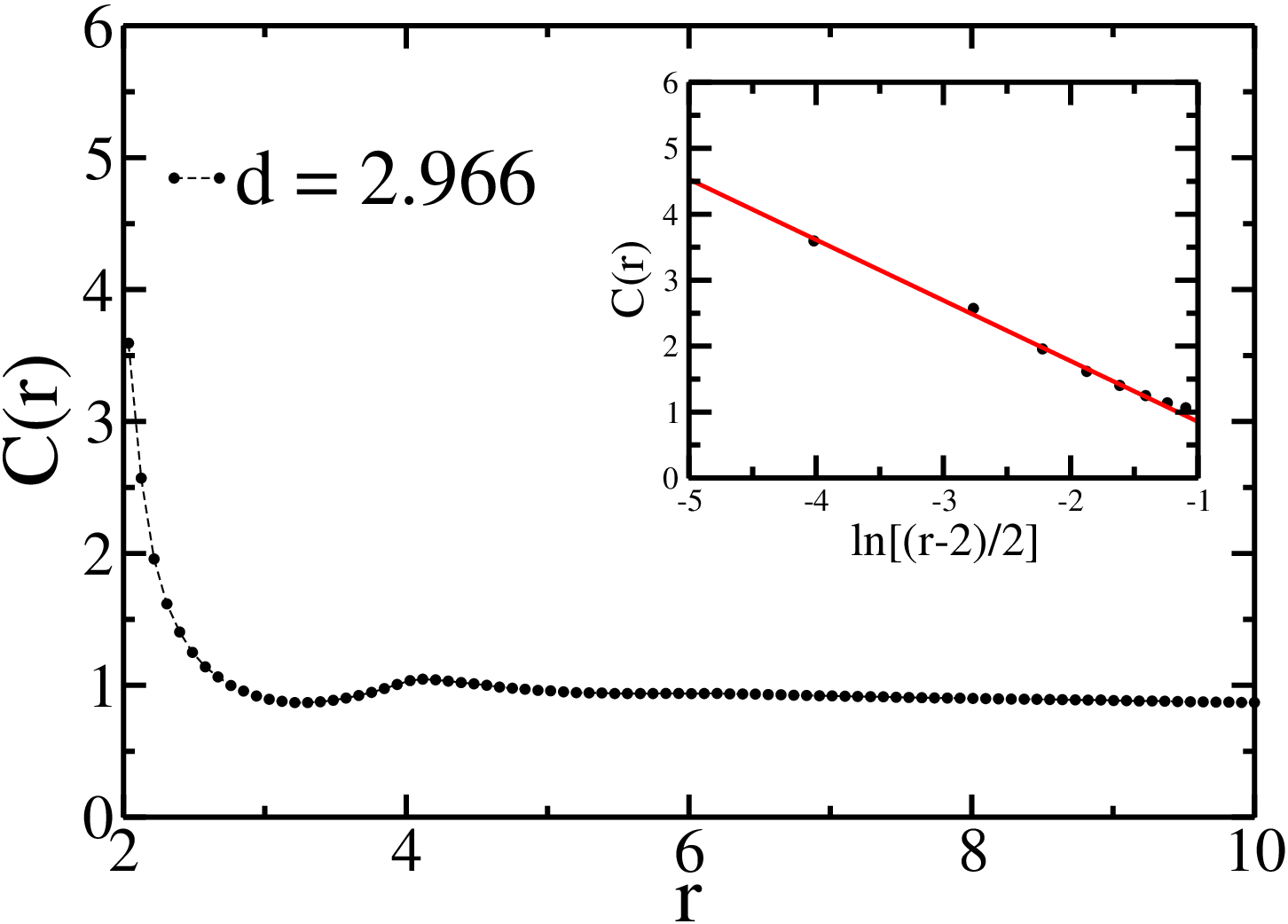}}
}
\caption{Two-point correlation function for collectors of different dimensions. Insets show asymptotic character for molecules in close proximity to each other with a fit given by (\ref{eq:corlog}). (a)~$d=1.4650$, (b)~$d=2.0$, (c)~$d=2.4022$, (d)~$d=2.5789$, (e)~$d=2.7268$ (Menger sponge), (f)~$d=2.8540$, (g)~$d=2.9656$}
\label{fig:cor}
\end{figure}

Two-point correlation functions for all fractal collectors studied are presented in Fig.\ref{fig:cor}.
For $d$ close to $3$, density autocorrelation function has got a typical shape. Its decay is fast and it has logarithmic singularity at kissing limit ($r \to 2 r_0^+$), exactly as predicted by (\ref{eq:corfast}) and (\ref{eq:corlog}). Situation changes in two cases in which $d \le 2$. Then, the correlations have long range structure. This is due to relatively sparse collectors which strongly limit space where spheres could be adsorbed. The same causes unpredictable properties at small distances, because they do depend mainly on a specific collector structure.
\section{Summary}
Random Sequential Adsorption on Menger sponge-like fractal collectors has been studied. For such systems, Feder's law (\ref{eq:feder}) slightly overestimates collector dimension. Analysis of two-point correlation function shows that, in general, it has logarithmic singularity at kissing limit and its decay seems to be superexponential. The only exceptions there are the sparsest collectors where internal structure overcomes any universal properties of adsorption. Obtained maximal random coverage ratios fit well  phenomenological relation proposed in \cite{bib:Torquato2006, bib:Ciesla2012b}.
\par
This work was supported by grant MNiSW/N N204 439040.
%


\begin{thebibliography}{10}
\bibitem{bib:Conway1998} J. H. Conway and N. J. A. Sloane, Sphere Packings, {\em Lattices and Groups}, Springer-Verlag, New York, 1998.
\bibitem{bib:Rogers1964} C. A. Rogers, {\em Packing and Covering}, Cambridge University Press, Cambridge, 1964.
\bibitem{bib:Hales2006} Int. J. Math. and Comp. Sci. {\bf 36} (1) 21 (2006).
\bibitem{bib:Renyi1963} A. Reńyi, Sel. Trans. Math. Stat. Prob. {\bf 4}, 203 (1963).
Publ. Math. Inst. Hung. Acad. Sci. {\bf 3}, 109, (1958).
\bibitem{bib:Privman2004} A.M.R. Cadilhe, V. Privman, Modern Phys. Lett. B {\bf 18}, 207 (2004).
\bibitem{bib:Fan1991} Y.Fan, J.K.Percus, Phys. Rev. Lett. {\bf 67} 13 1677 (1991).
\bibitem{bib:Feder1980} J. Feder, J. Theor. Biol. 87, 237 (1980).
\bibitem{bib:Swendsen1981} R. H. Swendsen, Phys. Rev. A {\bf 24}, 504 (1981).
\bibitem{bib:Privman1991} V. Privman, J.-S. Wang, and P. Nielaba, Phys. Rev. B {\bf 43} 3366 (1991).
\bibitem{bib:Torquato2006} S. Torquato, O.U. Uche, F.H. Stillinger, Phys.Rev.E {\bf 74} 061308 (2006).
\bibitem{bib:Torquato2002} S. Torquato and F. H. Stillinger, J. Phys. Chem. B {\bf 106}, 8354 (2002).
\bibitem{bib:Ciesla2012b} M. Ciesla, J. Barbasz, J. Chem. Phys. {\bf 137} 044706 (2012).
\bibitem{bib:Bonnier1994} B. Bonnier, D. Boyer, and P. Viot, J. Phys. A {\bf 27}, 3671 (1994).
\bibitem{bib:Fuchs2001} A.H. Fuchs, A.K. Cheetham, J. Phys. Chem. B,  {\bf 105} (31) 7375 (2001).
\bibitem{bib:Kinge2008} S. Kinge, M. Crego-Calama, D.N. Reinhoudt, ChemPhysChem
{\bf 9} 1 20 (2008).
\bibitem{bib:Pfeifer1983} P. Pfeifer and D. Avnir, J. Chem. Phys. {\bf 79}, 3558 (1983)
\bibitem{bib:Avnir1983} D. Avnir, D. Farin, and P. Pfeifer, J. Chem. Phys. {\bf 79}, 3566 (1983); Surf. Sci. {\bf 126}, 569 (1983); Nature {\bf 30}, 261 (1984).
\bibitem{bib:Basillais1998} E. Basillais, Comptes Rendus de l'Academie des Sciences - Serie III, {\bf 321} (4), 295 (1998).
\bibitem{bib:Khasanov1991} M. M. Khasanov and I. I. Abyzbaev, J. Eng. Phys. Thermophys. {\bf 61} 6, 1516 (1991).
\bibitem{bib:Nazzarro1996} M. S. Nazzarro, A. J. Ramirez Pastor, J. L. Riccardo and V. Pereyra, J. Phys. A: Math. Gen. {\bf 30} 1925 (1997).
\bibitem{bib:Loscar2003} E. S. Loscar, R. A. Borzi, and E. V. Albano, Phys. Rev. E {\bf 68} 041106 (2003).
\bibitem{bib:Cole1986} M. W. Cole and N. S. Holter, Phys. Rev. B {\bf 33} 8806 (1986).
\bibitem{bib:Pfeifer1989} P. Pfeifer, Y. J. Wu, M. W. Cole, J. Krim, Phys. Rev. Lett. {\bf 62} 1997 (1989).
\bibitem{bib:Brilliantov1998} N. V. Brilliantov, Yu. A. Andrienko, P. L. Krapivsky, and J. Kurths, Phys. Rev. E {\bf 58} 3530 (1998).
\bibitem{bib:Ciesla2012a} M. Ciesla, J. Barbasz, J. Stat. Mech. 03015 (2012).
\bibitem{bib:Adamczyk2010} Z. Adamczyk, J. Barbasz, M. Ciesla, Langmuir, {\bf 26} 11934 (2010);
\bibitem{bib:Adamczyk2011} Z. Adamczyk, J. Barbasz, M. Ciesla, Langmuir, {\bf 27} 6868 (2011).
\bibitem{bib:Tarjus1991} G. Tarjus, P. Schaaf, J. Talbot, J. Stat. Phys. {\bf 63} 167 (1991).
\bibitem{bib:AdamczykBook} Z. Adamczyk, {\em Particles at Interfaces: Interactions, Deposition, Structure}, Elsevier/Academic Press, Amsterdam, 2006.
\bibitem{bib:Ball1992} K. Ball, Int. Math. Res. Notices {\bf 68}, 217 (1992).
\end{thebibliography}
\end{document}